\documentclass[%
 reprint,
nofootinbib,
 amsmath,amssymb,
 aps,
 prl
]{revtex4-1}

\usepackage{aas_macro}
\usepackage{graphicx}
\usepackage{dcolumn}
\usepackage{bm}

\begin{document}


\title{Observed instability constraints on electron heat flux in the solar wind}


\author{Yuguang Tong}
\email{ygtong@berkeley.edu}
\author{Stuart D. Bale}
\affiliation{Physics Department, University of California, Berkeley, California, USA}
\affiliation{Space Sciences Laboratory, University of California, Berkeley, California, USA}

\author{Chadi Salem}
\author{Marc Pulupa}
\affiliation{Space Sciences Laboratory, University of California, Berkeley, California, USA}



\date{\today}

\begin{abstract}
We present for the first time the joint distribution of electron heat flux and electron plasma beta, and the joint distribution of core electron drift velocity and electron plasma beta using a statistically large dataset of solar wind electrons at 1AU. We calculate the growth rates of linear instabilities and show compelling experimental evidence that the heat-flux-driven Alfv{\'e}n instabilities constrain the electron core drift, and therefore the electron heat flux, in low-beta plasmas. This result is relevant to understanding energy transport in low-beta solar/stellar coronae and winds. 

\end{abstract}

\pacs{Valid PACS appear here}
\maketitle


\section{Introduction}
Energy transport via heat conduction is fundamental to fusion \cite{Hinton:1976a}, space \cite{Parker:1964a, DePontieu:2011a} and astrophysical \cite{Cowie:1977a} plasmas. The micro-physics of heat conduction in weakly collisional and magnetized plasmas is not well understood,  although it is significant for modeling global systems. Observations of heat flux (HF) in collisionless solar wind \cite{Scime:1994a, Bale:2013a} and intracluster medium (ICM) \cite{Markevitch:2000a, Markevitch:2003a} show significant departure from the classical Spitzer-H{\"a}rm HF, suggesting the necessity of HF regulation mechanisms in absence of Coulomb collisions.

In high-$\beta_e$ ($\beta_e \equiv 8\pi n_e k_B T_e / B_0^2$) space and astrophysical plasmas, HF constraint by whistler instabilities has been studied extensively by theories \cite{Gary:1975b, Gary:1994a, Pistinner:1998a, Gary:2000a} and simulations \cite{Roberg-Clark:2018a, Komarov:2017a}, predicting a marginal HF scaling $q_e\propto 1/\beta_e$, which is consistent with solar wind observations \cite{Gary:1999a}. However, the HF regulation mechanism in low-$\beta_e$ plasmas is less well understood due to scarcity of observational data \cite{Gary:1998b, Gary:1999a}. In this Letter, we fill the gaps by presenting the first joint distribution of HF and electron beta in the solar wind at 1AU using a statistically large dataset, and showing compelling experimental evidence that HF-driven Alfv{\'e}n instabilities effectively limit HF in the low-$\beta_e$ solar wind. 

The solar wind is a highly-ionized, weakly collisional plasma that expands super-Alfv{\'e}nically from the solar corona into the heliosphere. Significant field aligned electron HF arises from the anisotropy in the electron velocity distribution function (eVDF), which is often modeled as a superposition of three gyrotropic populations: a cool dense `core', a hotter tenuous `halo' \cite{Feldman:1975a}, and a beam-like field-aligned `strahl' \cite{Rosenbauer:1977a}. In the solar wind frame (here, we define the solar wind frame as the frame in which the total current of all solar wind ion species is zero), the core electrons drift sunward along the background magnetic field line, whereas the suprathermal (halo and strahl) electrons drift anti-sunward. The drift velocities of core and suprathermal electron populations are anti-correlated, maintaining near zero net electric current in the solar wind frame \cite{Feldman:1975a, Scime:1994a}.

\section{Solar wind electron measurements}
\label{section:dataset}
 We use a solar wind electron dataset \cite{Pulupa:2014} produced by non-linear fits to the electron velocity distribution function (VDF), measured by two electron detectors (EESA-L and EESA-H \cite{Lin:1995a}) onboard the \emph{Wind} spacecraft. Spacecraft potential is corrected using independent measurements of electron density by the \emph{Wind}/WAVES Thermal Noise Receiver (TNR) \cite{Bougeret:1995a} using quasi-thermal noise analysis \cite{Meyer-Vernet:1989a}. 

Each electron VDF is fitted to a two-component model:
$f_{fit}(v_{||}, v_\perp) = f_c(v_{||}, v_\perp) + f_h(v_{||}, v_\perp)$
where $f_c$ and $f_h$ correspond to the core and halo electrons. The core distribution takes the form of a drifting, bi-Maxwellian
$f_c(v_{||}, v_\perp) = A_c\exp{\left[-\left(v_{||}-v_{dc}\right)^2/v_{Tc||}^2 - v_\perp^2/v_{Tc\perp}^2\right]}$, where $A_c = 2n_c v_\perp /\sqrt{\pi} v_{Tc\perp}^2v_{Tc||}$ and $v_{Tcj} = \sqrt{2T_{cj}/m_e}$, $(j=||, \perp)$. The halo distribution takes the form of a drifting, bi-kappa
$f_h(v_{||}, v_\perp) = A_h B_h$ $\left[1+\left(v_{||}-v_{dc}\right)^2/v_{Th||}^2 + v_\perp^2/v_{Th\perp}^2\right]^{-\kappa-1}$
where $A_h=2n_h v_\perp /\sqrt{\pi}v_{Th\perp}^2v_{Th||}$, $B_h=\Gamma(\kappa+1)/\Gamma(\kappa-1/2)$ and the thermal velocities $v_{Thj} = \sqrt{(2\kappa-3)T_{hj}/m_e}$, $(j=||, \perp)$. 

The strahl electrons are characterized by the difference between the measured VDF and the fitted VDF: $f_s(v_{||}, v_\perp)=f(v_{||}, v_\perp)-f_{fit}(v_{||}, v_\perp)$. In this Letter we use the strahl density $n_s$ and the effective drift speed $v_{ds}$, as computed from the zeroth and first moments of $f_s$.

\begin{figure*}
\begin{center}
\includegraphics[width=0.9\textwidth]{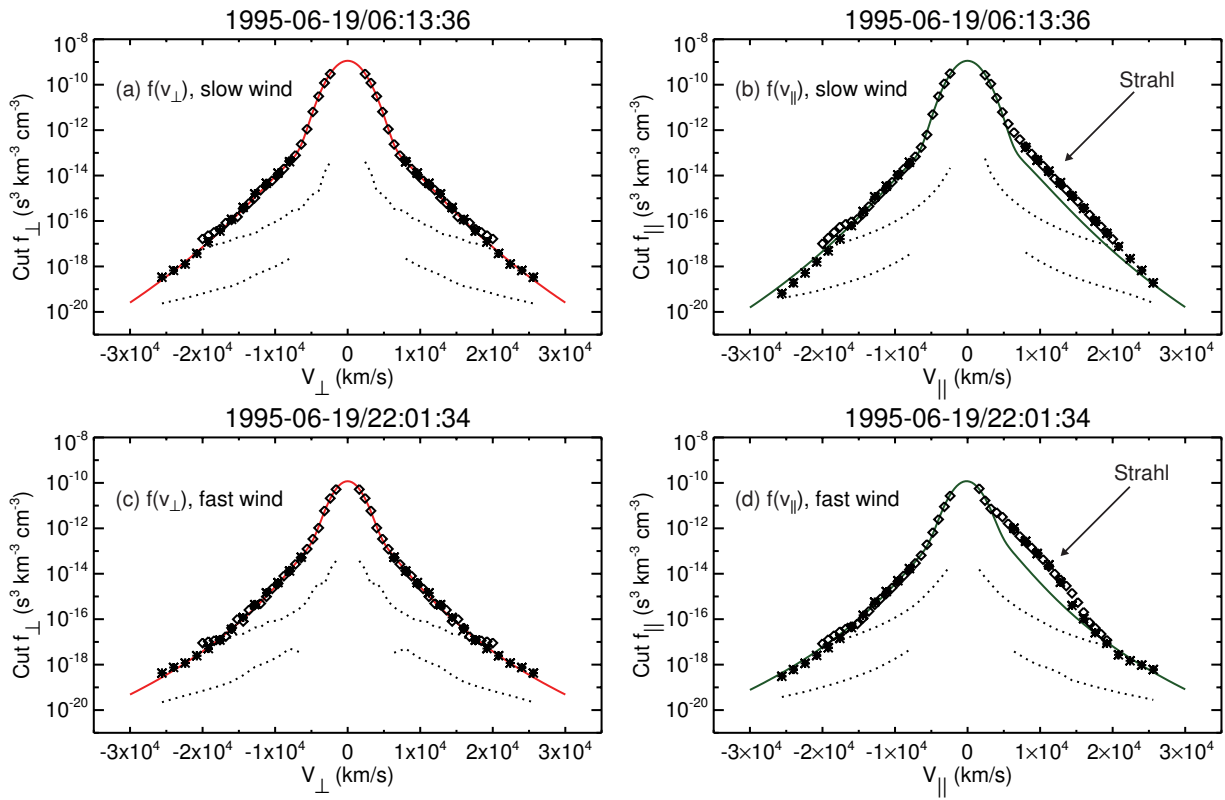}
\caption{
Measured solar wind electron distribution functions and fits. The top two panels show $f(v)$ in the slow solar wind ($v_{sw} \approx$ 381 km/s), with $f(v_\perp)$ in panel (a) and $f(v_{\parallel})$ in panel (b).  Hollow squares show the EESA-L data, while asterisks mark the EESA-H data and small black dots mark the 1-count level for each detector. The red curve in panel (a) is the fit to the measured $f(v_\perp)$.  The green curve
in panel (b) is the fit to $f(v_{\parallel})$. The strahl appears clearly as an enhanced field-aligned feature which is limited in energy.
Panels (c) and (d) show the same features in the fast solar wind ($v_{sw} \approx$ 696 km/s).}
\label{fig:fit}
\end{center}
\end{figure*}

Figure \ref{fig:fit} shows typical electron VDFs measured in the fast and slow solar wind; fast wind is characterized by hotter,  more tenuous plasma than slow wind.  The top panels show cuts through the perpendicular velocity distribution $f(v_\perp)$ and the parallel velocity distribution $f(v_\parallel)$ on the left and right, respectively, for an interval of slow solar wind.  Data from EESA-L and EESA-H are included, shown as diamonds and stars respectively, and the `one count' levels for each detector are shown as dotted black lines.  The two detectors are well inter-calibrated.  The bottom panels show the same for an interval of fast wind. 

Following the aforementioned procedure, we obtain 154,567 independent electron measurements from two, 2 year intervals: 1995–-1997 (solar minimum) and 2001-–2002 (solar maximum), and include only `ambient' solar wind (no coronal mass ejections (CMEs), foreshock, etc.). Intervals of `bidirectional' heat flux (usually associated with CMEs) are also excluded. We define the direction of strahl electron HF to be the positive direction, hence the core electron drift is negative within fitting error. 
For a single nonlinear fit, the fitted core electron drift velocity is comparable to the fitting error \cite{Pulupa:2014}. We invoke the zero current condition to infer core electron drifts from densities and drifts of suprathermal electrons \cite{Feldman:1976a, Feldman:1976b, Scime:1994a}: $v_{dc} = -(n_h v_{dh} + n_s v_{ds})/n_c$.

\section{Linear theory}
\label{section:linearTheory}
Growth rate contours of linear instabilities have been proven successful  in characterizing marginally stable plasmas \cite{Hellinger:2006a}. Linear Alfv{\'e}n HF instabilities have been studied in the literature \cite{Gary:1975b, Gary:1998b, Tong:2015c} by solving the full set of the linear Vlasov-Maxwell equations. We extend the above calculation to cover the range $\beta_{c\parallel} \in [0.1, 1]$. The electron VDF is modeled by two drifting bi-Maxwellian components representing the core and halo, while the proton VDF is assumed to be isotropic and Maxwellian. We also assume charge neutrality $n_p=n_c+n_h$ and zero current $n_cv_{dc}+n_hv_{dh}=0$.

Simple wave-particle resonance analysis shows that the Alfv{\'e}n HF instability strongly Landau resonates with the core electrons, weakly Landau/cyclotron resonates with protons, and barely interacts with the high energy halo and strahl electrons \cite{Gary:1975b, Gary:1998b}. Therefore, ignoring the strahl is a reasonable assumption for the growth rate calculation and should not limit our discussion to the slow solar wind where strahl electrons are scarce. In fact we shall see that the linear theory and the electron data agrees well in low-$\beta_{c||}$ solar wind, which is often fast wind. The same argument based on resonance analysis justifies our usage of bi-Maxwellian VDF for halo electrons. Using a drifting kappa VDF will be more accurate but we expect that the correction is small \cite{Saeed:2017a}. 

We use the following typical solar wind dimensionless parameters (which are consistent with the dataset): $n_c/n_p = 1-n_h/n_p = 0.95$,  $v_A/c = 10^{-4}$, $T_{c||}/T_{c\perp} = T_{h||}/T_{h\perp}=1$ and $T_{h||}/T_{c||} = 6$. The three variables are $v_{dc}/v_A$, $\beta_{c||}$ and $T_{c}/T_p$. For every point in the $(-v_{dc}/v_A, \beta_{c||})$ plane, we search the wavenumber space for the maximum growth rate of the Alfv{\'e}n HF instabilities with $T_c/T_p$ taking value of $0.5, 1$ or $2$. 

Whistler HF instabilities have been studied in linear theories \cite{Gary:1975b, Gary:1994a, Gary:1999a}, quasi-linear theory \cite{Pistinner:1998a} and nonlinear kinetic simulations \cite{Roberg-Clark:2018a, Komarov:2017a}, all showing the similar scaling $q_e\propto 1/\beta_e$. In this Letter, we use the linear whistler HF instability threshold on the electron HF at a fixed growth rate $\gamma/\Omega_{p} = 10^{-2}$ from \cite{Gary:1999a}:

\begin{equation}
\frac{q_e}{q_{0}} = \frac{1.47}{\sqrt{2}\beta_{c\parallel}^{1.07}}
\label{eq:gary1999formula}
\end{equation}
where $q_{0} = 3/2n_e k_B T_{c\parallel} v_{Tc||}$ is the free streaming heat flux corresponding to transport of the thermal energy at the thermal speed\footnote{Note that \cite{Gary:1999a} defines $v_{Tc}= \sqrt{k_BT_{c\parallel}/m_e}$ instead, leading to a difference of $\sqrt{2}$ in Eq. (\ref{eq:gary1999formula}).}.

In this simple drifting core-halo model, electron drift and electron HF are linearly correlated in the absence of temperature anisotropy \cite{Feldman:1976a}:

\begin{equation}
\frac{q_e}{q_{0}} = \frac{5}{3} \frac{n_c}{n_e} \frac{v_{dc}}{v_{Tc}}\left(\frac{T_{h\parallel}}{T_{c\parallel}}-1\right)
\label{eq:twoComponentQ}
\end{equation}

\section{Instability constraints on core drifts and electron heat flux}

\begin{figure}
\begin{center}
\includegraphics[width=0.45\textwidth]{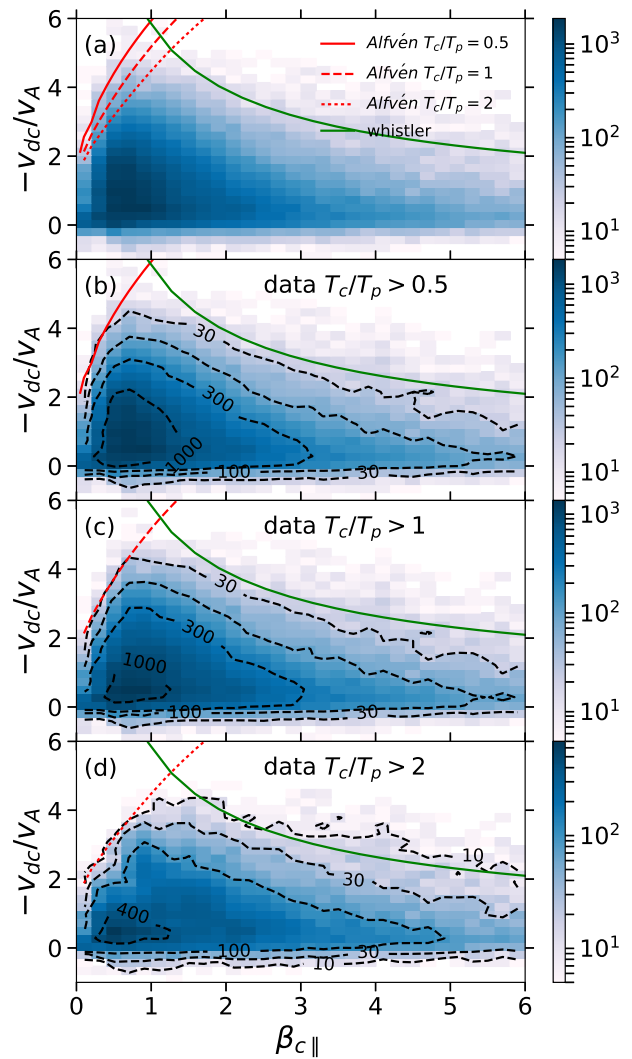}
\caption{Joint distributions of the dimensionless core electron drift speed and the core electron parallel beta with data contours in black dashed curves. Overlaid color lines represent the $\gamma/\Omega_p=10^{-2}$ linear growth rate coutours associated with Alfv{\'e}n waves (red) and whistler waves (green). The joint distribution in panel (a) uses all data in the dataset, while those in panel (b)-(d) are conditioned on different thresholding core-proton-temperature ratios.}
\label{fig:driftBetaPanel}
\end{center}
\end{figure}

Panel (a) of Figure \ref{fig:driftBetaPanel} shows the joint distribution of $-v_{dc}/v_A$ and $\beta_{c||}$. The overplotted color curves show the contours of the maximum growth rates at $\gamma/\Omega_p=10^{-2}$ in the core-halo-proton plasmas with  $-v_{dc}/v_A$ and $\beta_{c||}$: the three red curves correspond to the Alfv{\'e}n HF instabilities with $T_c/T_p = 0.5$ (solid), $1$ (dashed) and $2$ (dotted); the green curve stands for the whistler HF instability given by Eq. (\ref{eq:twoComponentQ}) and Eq. (\ref{eq:gary1999formula}). Other plasma parameters are given in the previous section. There is an apparent upper bound on $-v_{dc}/v_A$ that increases with $\beta_{c||}$ until $\beta_{c||}\sim 1$ and then decreases with $\beta_{c||}$. This feature is consistent with the Alfv{\'e}n HF instability threshold in the low-$\beta_{c||}$ regime and with the whistler HF instability threshold in the higher-$\beta_{c||}$ regime. The differences among the red curves in the panel show that the Alfv{\'e}n HF instability is highly susceptible to the core-proton temperature ratio $T_c/T_p$, since the instability is resonant with both core electrons and protons \cite{Gary:1998b}. 

Panels (b)-(d) of Figure \ref{fig:driftBetaPanel} show the same joint distribution $(-v_{dc}/v_A, \beta_{c||})$ as the panel (a) but with conditions $T_c/T_p>0.5, 1, 2$ respectively. The Alfv{\'e}n HF instability growth rate contours based on the linear kinetic theory align strikingly well with the empirical joint distributions contours at different values of $T_c/T_p$. We interpret the above agreement as compelling evidence that the Alfv{\'e}n HF instability regulates the core electron drifts in low$-\beta_{c||}$ solar wind at $1$AU.

\begin{figure}
\begin{center}
\includegraphics[width=0.45\textwidth]{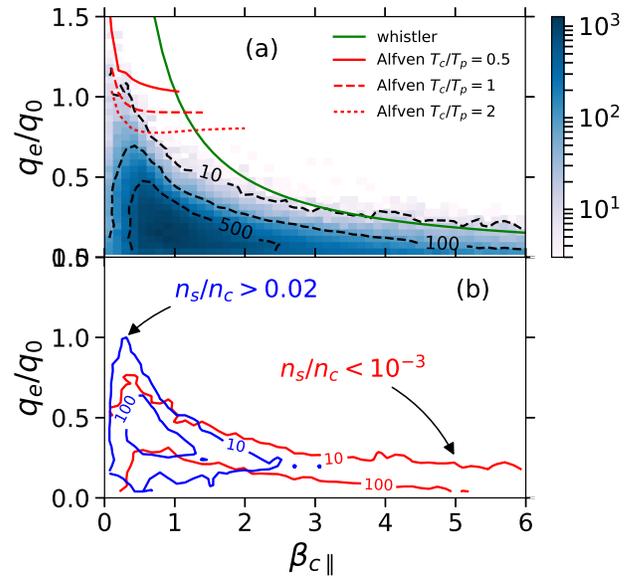}
\caption{The upper panel shows the joint distribution of the normalized electron HF and the core electron parallel beta overplotted with linear wave instability contours (red for Alfv{\'e}n waves and green for whistler waves) at $\gamma/\Omega_p=10^{-2}$.  The lower panel shows the data contours when thresholding for abundant (blue) and scarce (red) strahl electrons.}
\label{fig:heatfluxBetaPanel}
\end{center}
\end{figure}

We also show instability constraints on the electron HF. The top panel of Figure \ref{fig:heatfluxBetaPanel} shows color scale plot of the joint distribution of $q_e/q_0$ and $\beta_{c||}$, overplotted with the instability thresholds at $\gamma /\Omega_p = 10^{-2}$. The green curve shows the whistler HF instability threshold corresponding to Eq. (\ref{eq:gary1999formula}). The three red curves are the theoretical Alfv{\'e}n HF instability upper bounds on $q_e/q_0$ obtained by transforming the core electron drift thresholds (also red curves) in Figure \ref{fig:driftBetaPanel} using Eq. (\ref{eq:twoComponentQ}).  The empirical upper bound on $q_e/q_0$ show the similar dependence on $T_c/T_p$ as $-v_{dc}/v_A$ (not shown here) when $\beta_{c||} \lesssim 1$. Clearly the Alfv{\'e}n HF instability and the whistler HF instability effectively impose an upper bound on the electron HF in observation. 

\begin{figure}
\begin{center}
\includegraphics[width=0.45\textwidth]{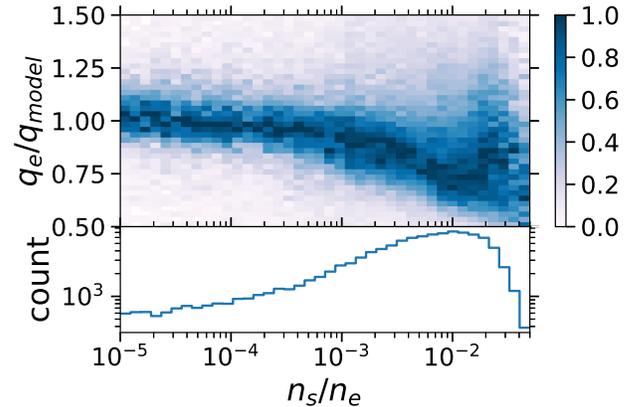}
\caption{Joint normalized distribution of $q_e/q_{model}$ and $n_s/n_e$ and the distribution of data with $n_s/n_e$. $q_{model}$ is the electron HF calculated from core drift by assuming the two-component-electron model (Eq. (\ref{eq:twoComponentQ})). }
\label{fig:strahlHeatflux}
\end{center}
\end{figure}

We notice that while the theoretical Alfv{\'e}n HF instability threshold stays above the observed values of $q_e/q_0$, the constraint is not as tight as in the case of core electron drifts (c.f. Figure \ref{fig:driftBetaPanel}). Also in comparison, the whistler HF instability threshold imposes a tighter constraint on $q_e/q_0$ for $\beta_{c||} \gtrsim 2$. The tightness of the theoretical bounds on $q_e/q_0$ can be shown to be related to the abundance of strahl electrons. Since the electron VDF in our linear theory does not include strahl electrons, the theoretical relation between core electron drifts and electron HF (Eq. (\ref{eq:twoComponentQ})) departs slightly from observations. Figure \ref{fig:strahlHeatflux} shows the joint normalized distribution of $q_e/q_{model}$ and $n_s/n_e$, where $q_e$ is the observed electron HF, and $q_{model}$ is the HF calculated from Eq. (\ref{eq:twoComponentQ}) using the core electron drift. When the fraction of strahl electrons is low, the model HF agrees well with observations, yielding $q_e/q_{model}\approx 1$. However, as the fraction of strahl electron increases, the two-component model often overestimates the HF at a given core electron drift. We estimate that the presence of strahl electrons could lead to an order unity correction to Eq. (\ref{eq:twoComponentQ}).

Returning to Figure \ref{fig:heatfluxBetaPanel},  the bottom panel shows the contours of the empirical joint distribution of $q_e/q_0$ and $\beta_{c||}$ conditioned on low strahl fraction (blue, $n_s/n_c>0.02$) and high strahl fraction (red, $n_s/n_c< 10^{-3}$ ). Apparently the occurrence of high fraction of strahl electrons concentrates in the low-$\beta_{c||}$ regime, justifying that the Alfv{\'e}n HF instability threshold on the electron HF is not as tight as that on the core electron drift.

\section{Discussion}
We show that in low-$\beta_{c||}$ solar wind the Alfv{\'e}n HF instability effectively constrains the sunward core electron drift velocity $v_{dc}$ and therefore also constrains the electron HF $q_e$ by the strong correlation between $v_{dc}$ and $q_e$, which is linear (Eq. (\ref{eq:twoComponentQ})) if temperature anisotropy and strahl electrons are ignored. Our result suggests that low-beta solar wind has an upper bound on HF which directly depends on the local plasma properties, differing from the classical Spitzer-H{\"a}rm HF and the free streaming HF. This is relevant to modeling the energy transport and the dynamics of the coronae and winds associated with our sun and other stars.

We note that the dataset of solar wind electrons at 1AU is dominated by plasmas with $\beta_{c||}\sim O(1)$ (see Figure \ref{fig:driftBetaPanel}). Since the plasma beta decreases with closer distance to the sun, coming missions exploring the inner heliosphere, namely the Parker Solar Probe and the Solar Orbiter will further test the role of the Alfv{\'e}n HF instability in low-beta space plasmas.  

We also note that the Alfv{\'e}n HF instability generates sunward-propagating kinetic Alfv{\'e}n waves at $k\rho_p \sim 1$ with $\theta_{kB}\sim 80^\circ$, providing a means to generate kinetic scale fluctuations without explicitly invoking nonlinear turbulence cascade. Enhanced field fluctuations may also modify core electron and proton heating.

The authors acknowledge helpful discusisions with Alfred Mallet, Trevor Bowen, David Sundkvist and Ivan Vosko. Work on Wind/3DP data at UC Berkeley is supported in part by NASA grant NNX16AP95G, and work on Solar Wind Electrons is supported by NASA Grant NNX16AI59G and NSF SHINE 1622498.

%


\end{document}